%% file: TOP2021proceedings_Negro.tex
\documentclass[12pt]{article}
\usepackage{graphicx}
\usepackage{url}

\newcommand\pubnumber{CMS CR-2022/002}
\newcommand\pubdate{January 17, 2022}

\def\institute{$^1$Purdue University}
\def\Title#1{\begin{center} {\Large #1 } \end{center}}
\def\Author#1{\begin{center}{ \sc #1} \end{center}}
\def\Address#1{\begin{center}{ \it #1} \end{center}}

\newcommand\pubblock{\rightline{\begin{tabular}{l} \pubnumber\\
         \pubdate  \end{tabular}}}
\newenvironment{Abstract}{\begin{quotation}  }{\end{quotation}}
\newenvironment{Presented}{\begin{quotation} \begin{center} 
             PRESENTED AT\end{center}\bigskip 
      \begin{center}\begin{large}}{\end{large}\end{center} \end{quotation}}

\input econfmacros.tex
\begin{document}
\begin{titlepage}
\pubblock

\vfill
\Title{Top Quark Modelling and Tuning in ATLAS and CMS}
\vfill
\Author{ Giulia Negro$^1$\\
on behalf of the ATLAS and CMS Collaborations} 
\Address{\institute}
\vfill
\begin{Abstract}
Recent results on top quark modelling and tuning are presented. 
In particular, the focus of this talk is on the studies on the definition and commissioning of the common Monte Carlo effort carried out by ATLAS and CMS within the LHCtopWG. 
An overview of current recommendations for modelling uncertainties is also presented. 
\end{Abstract}
\vfill
\begin{Presented}
$14^\mathrm{th}$ International Workshop on Top Quark Physics\\
(videoconference), 13--17 September, 2021
\end{Presented}
\vfill
\end{titlepage}
\def\thefootnote{\fnsymbol{footnote}}
\setcounter{footnote}{0}

\section{Introduction}
One of the crucial ingredients in top quark analyses is the Monte Carlo (MC) simulation, because good modelling of data and high accuracy predictions are needed in order to make interpretations. 
Well-defined and possibly small uncertainties are also needed, since these are a limiting factor in many precision measurements and searches. 
In ATLAS~\cite{ATLAS:2008xda} and CMS~\cite{CMS:2008xjf} the same kind of generators is used but the prescriptions for modelling uncertainties are different, so their effects are sometimes hard to compare. 
However, if we want accurate combinations, it is critical to understand how to combine the differing strategies of ATLAS and CMS and to discuss how to reduce modelling systematics.

\section{Overview of current recommendations for modelling uncertainties}
Both ATLAS and CMS have a list of \emph{standard} recommendations to assess the modelling uncertainties, which should, however, not be considered final. \\ 
The matrix-element (ME) uncertainties are addressed with variations of the renormalization ($\mu_R^\mathrm{ME}$) and factorization ($\mu_F^\mathrm{ME}$) scales. \\
For the PDF uncertainties, the PDF4LHC recommendations are usually followed. \\
The top quark $p_\mathrm{T}$ modelling and top quark mass uncertainties are very analysis-dependent and will not be discussed here, while the uncertainties involving parton shower (PS) generators will be described in the following. \\
For the uncertainty related to initial (ISR) and final (FSR) state radiation, both experiments use independent $\mu_R^\mathrm{ISR}$ and $\mu_R^\mathrm{FSR}$ scale variations with factors (2, 0.5). \\
The underlying event (UE) uncertainty is addressed by both experiments considering variations of their nominal tunes, A14 for ATLAS and CP5 for CMS. \\
For the color reconnection (CR) uncertainty, both ATLAS and CMS retune UE data with different CR models implemented in Pythia8. 
These models show an overall agreement between data and MC but the CR uncertainty still continues to be one of the dominant uncertainty sources on the most precise top quark mass measurement. \\ 
Heavy-quark fragmentation is described in Pythia by the Bowler-Lund fragmentation function. To address the b-quark fragmentation uncertainty both experiments perform variations of the $r_\mathrm{B}$ parameter of this function. 
In CMS, a reweighting at generator level is performed via a transfer function based on LEP data, and also a comparison to the Peterson fragmentation function is included, while in ATLAS dedicated MC samples with an $r_\mathrm{B}$ value from LEP data are used. 
A dedicated $\mathrm{t\bar{t}}$ measurement has been performed in CMS~\cite{CMS:2021sln}, 
where $r_\mathrm{B}$ is extracted from a template fit to fragmentation proxy distributions. 
The results are in agreement with those obtained at the Z boson mass pole in LEP data and a significant improvement in experimental precision is achieved. Also, the $r_\mathrm{B}$ value obtained is in good agreement with the value used in the Pythia8 function. 
A dedicated $\mathrm{t\bar{t}}$ measurement of fragmentation observables~\cite{ATLAS:2020iyi} has been performed also in ATLAS. 
Different MC generators (Powheg+Pythia, Powheg+Herwig, Sherpa) successfully predict the shape of the observables in data. \\
Fragmentation and hadronization, which are important sources of uncertainties in most mass measurements, are evaluated in both experiments comparing Pythia vs Herwig. 
In CMS, only the impact on the jet energy response is considered. \\
For the hadron decays uncertainty, both experiments vary the semi-leptonic branching fractions of B hadrons within their PDG uncertainties. \\
It is also important to validate the nominal prediction with independent samples. 
Both experiments compare Powheg to MC@NLO samples to address the uncertainty related to the generator-NLO matching scheme. 
In ATLAS this is included as an additional “matching” uncertainty, while in CMS it is used as a cross-check. \\ 
To evaluate the ME-PS matching uncertainty, both experiments consider variations of the $h_\mathrm{damp}$ parameter, which regulates the first high-$p_\mathrm{T}$ emission. 
In CMS, the $h_\mathrm{damp}$-related uncertainties are estimated from a fit to the leading additional jet $p_\mathrm{T}$, while in ATLAS the $h_\mathrm{damp}$ value is based on data but not fitted.  \\
Many of these uncertainties could benefit from a common ATLAS-CMS MC sample, since comparing the CMS vs ATLAS effect of these uncertainties using a common sample would help in understanding them better.

\section{A first $\mathrm{t\bar{t}}$  common sample}
A $\mathrm{t\bar{t}}$ MC sample with Common Settings could facilitate combinations and comparisons between ATLAS and CMS. 
It could help to understand correlations of systematic uncertainties due to MC modelling, and it could remove differences in high-precision measurements. 
It could also be used as a baseline prediction in differential distributions and it would be a first step towards sharing the resources between the experiments, for both current and future generators. 

First of all, the samples used in ATLAS and CMS were investigated. 
Both experiments use a similar setup for $\mathrm{t\bar{t}}$ simulation, POWHEG-BOX (hvq) + Pythia8, but their nominal samples are different because many parameters are different, as can be seen in Table~\ref{tab:highlights}. 
For the first proposal of Common Settings, a “democratic” setup (not optimized to data) was decided: the same Pythia \emph{Monash} tune (basis of both ATLAS and CMS tunes), approximate averages for all physical parameters, technical parameters mainly chosen from the ATLAS setup. 
\begin{table}[!h!tbp]
\begin{center}
\caption{Main POWHEG and Pythia8 settings used in the ATLAS and CMS default MC event generation setups for $\mathrm{t\bar{t}}$ production and proposal for Common Settings.~\cite{ATLAS:2021jgj}}
\resizebox{\textwidth}{!}{
\begin{tabular}{ l l c c | c}
\hline
    Setting name & Setting description & CMS default & ATLAS default & Common Proposal \\
    \hline
    \hline 
     \textbf{POWHEG} & & & & \\
    \hline 
    qmass        & top-quark mass   [GeV]   & 172.5       & 172.5  & 172.5    \\
    twidth       & top-quark width  [GeV]   & 1.31        & 1.32 & 1.315     \\
    hdamp        & first emission damping parameter [GeV] & 237.8775      & 258.75 & 250  \\
    \hline
    wmass        & $W^{\pm}$ mass   [GeV]   & 80.4        & 80.3999 & 80.4  \\ 
    wwidth       & $W^{\pm}$ width  [GeV]   & 2.141       & 2.085  & 2.11  \\
    bmass        & $b$-quark mass  [GeV]    & 4.8         & 4.95  & 4.875   \\
    \hline
    \hline 
      \textbf{PYTHIA 8} & & & & \\
  \hline
     & PYTHIA 8 version & v240 & v230 & v240 (CMS) \\
     &  &  &  & v244 (ATLAS) \\ 
                 & Tune & CP5 & A14 & Monash \\
    \hline 
    PDF:pSet & LHAPDF6 parton densities to be used for proton beams & NNPDF31\_nnlo & NNPDF23\_lo & NNPDF23\_lo \\
   & & \_as\_0118 & \_as\_0130\_qed & \_as\_0130\_qed \\
    TimeShower:alphaSvalue & Value of $\alpha_s$ at $Z$ mass scale for Final State Radiation & 0.118    & 0.127 & 0.1365  \\ 
    SpaceShower:alphaSvalue & Value of $\alpha_s$ at $Z$ mass scale for Initial State Radiation & 0.118    & 0.127 & 0.1365  \\ 
    MPI:alphaSvalue  & Value of $\alpha_s$ at $Z$ mass scale for Multi-Parton Interaction  & 0.118 & 0.126  & 0.130  \\
    MPI:pT0ref       & Reference $p_\mathrm{T}$ scale for regularizing soft QCD emissions     & 1.41    & 2.09   & 2.28  \\ 
    ColourReconnection:range   &  Parameter controlling colour reconnection probability   & 5.176       & 1.71       & 1.80  \\
\hline
\end{tabular}
}
\label{tab:highlights}
\end{center}
\end{table}

\begin{figure}[!h!tbp]
\centering
\includegraphics[height=2.6in]{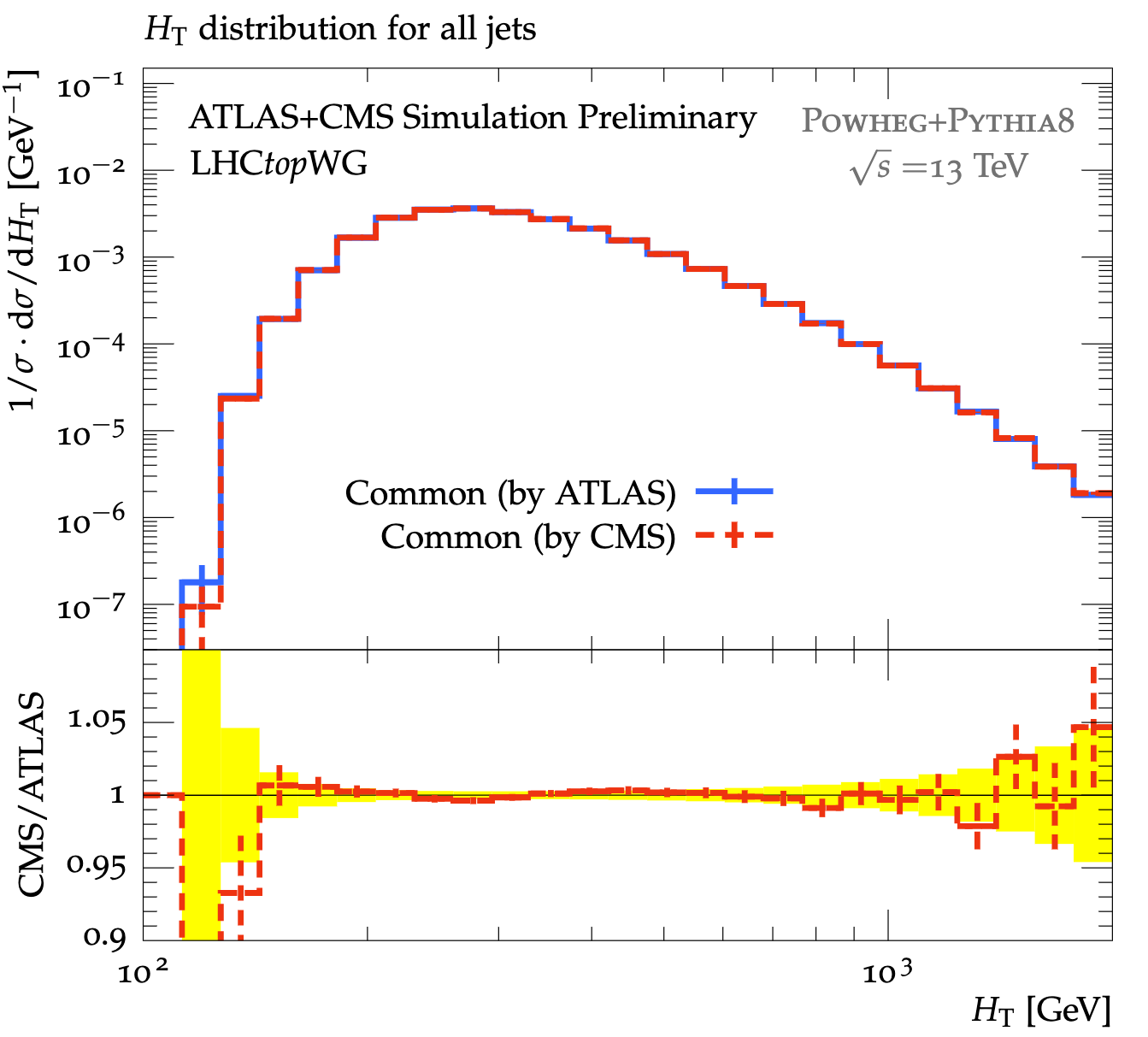}
\includegraphics[height=2.6in]{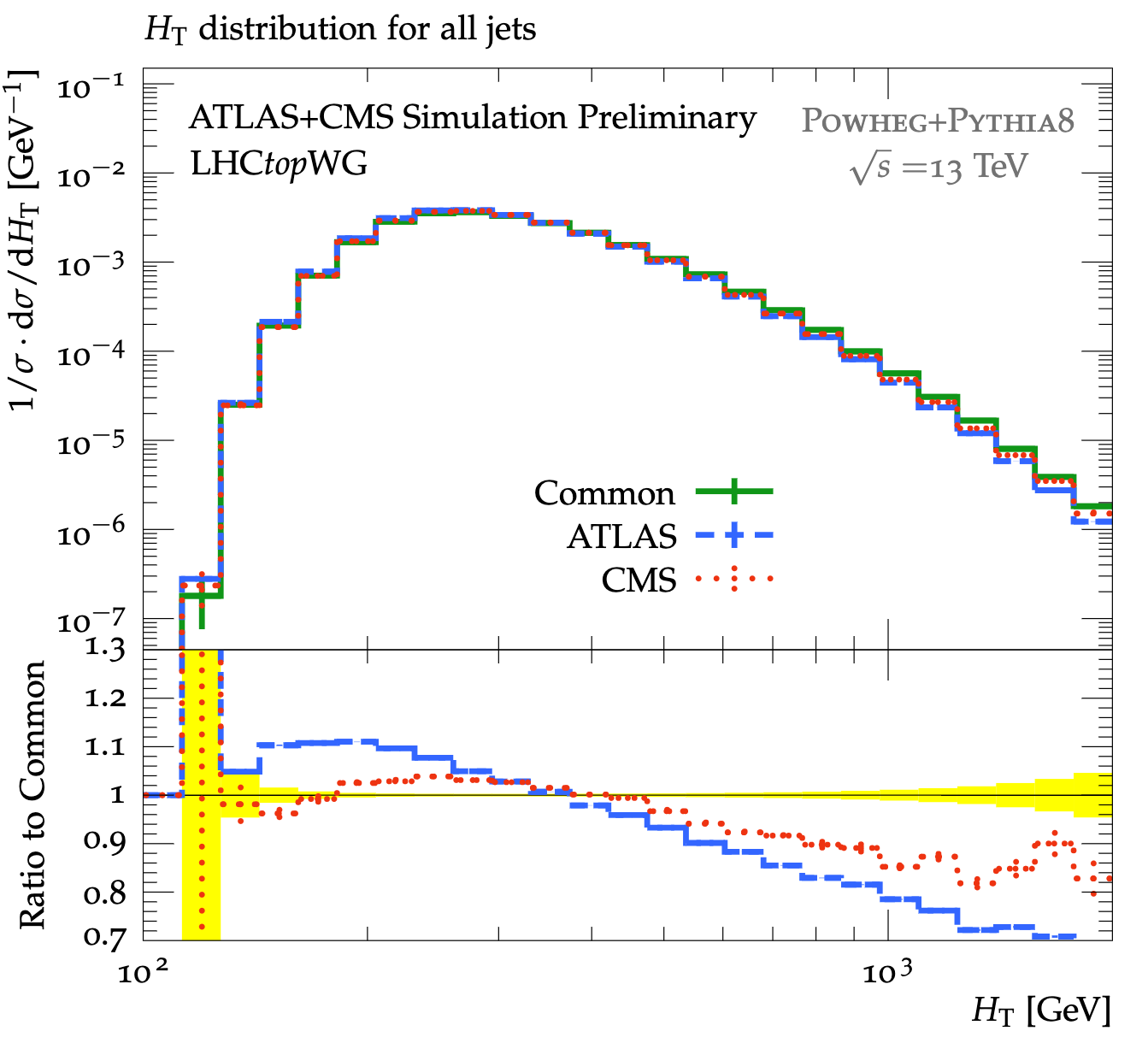}
\caption{[Left] Comparison of $\mathrm{t\bar{t}}$ MC samples produced by ATLAS (in blue) and CMS (in red) with Common Settings, generated with Powheg+Pythia8. 
[Right] Comparison of $\mathrm{t\bar{t}}$ MC samples produced with Common Settings (in green), and with nominal ATLAS (in blue) and CMS (in red) settings, generated with Powheg+Pythia8.~\cite{ATLAS:2021jgj} }
\label{fig:HT}
\end{figure}

After exchanging the complete set of parameters, the first sample with Common Settings (\emph{v0.1}) was produced independently in each collaboration. 
LHE files with 10M inclusive events were produced and showered separately in the respective frameworks, with a different Powheg revision but the same hvq program, a different Pythia version (that was checked to give identical results), and no use of EvtGen. \\
In order to validate the samples produced in the two experiments, 
comparisons between the samples were performed at particle level with Rivet v3.1.2 and the standard “MC\_TTBAR” routine, applying a one-lepton filter with the “ONELEP” mode. 
The distributions are in perfect agreement within statistical uncertainties, as shown on the left in Figure~\ref{fig:HT}. 
The common sample was then compared with the nominal samples of the two experiments, as shown on the right in Figure~\ref{fig:HT}. 
Few differences in the distributions can be seen, mainly due to the different $\alpha_S$ of the tune. 
Also, the ATLAS and CMS nominal samples are tuned to their experimental results, while the Common Settings are not optimized to data. 

In order to produce a first “physical” common sample (\emph{v0.2}) with Common Settings more tuned on data, Powheg and Pythia settings were agreed between ATLAS and CMS experts. 
In this iteration, common LHE files are used, produced by ATLAS and showered separately by CMS and ATLAS. 
For POWHEG settings, the values in \emph{v0.1} were mainly averaged between ATLAS and CMS, while now a choice more justifiable from the physics point of view was taken, with values from PDG or theory calculations. 
For Pythia8, the main “parameter” to be defined is the tune. 
In \emph{v0.1}, the settings from the \emph{Monash} tune were used, while now shower settings consistent with the \emph{Powheg Sudakov} are used together with the \emph{Monash} tune. \\
After deciding the new Common Settings (\emph{v0.2}) a few sets of events were produced and first comparisons with the old settings and the ATLAS and CMS data were performed.
Much better agreement between the common sample and the nominal samples from ATLAS and CMS is now achieved. 
The new settings and results will be documented in a new public note, including comparisons to data at parton and particle level with additional Rivet routines.

\section{A $\mathrm{t\bar{t}}$ sample with off-shell effects}
\begin{figure}[!h!tbp]
\centering
\includegraphics[height=2.5in]{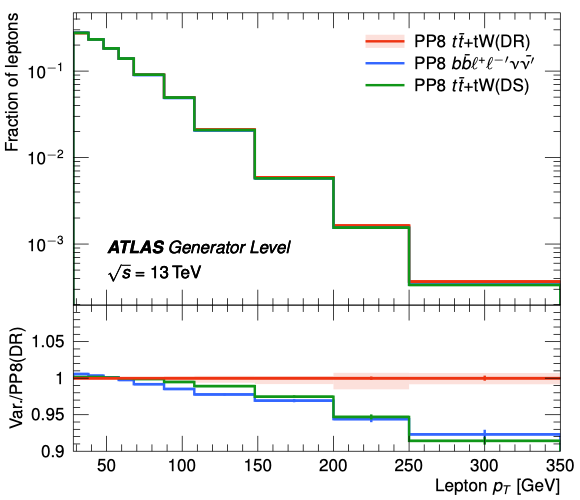}
\includegraphics[height=2.5in]{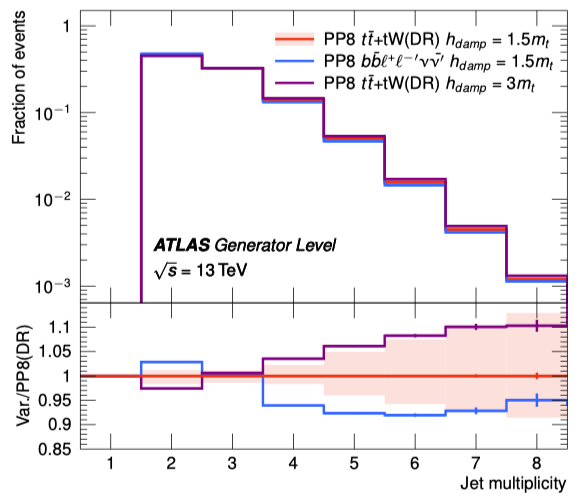}
\caption{[Left] Comparison of the standard $\mathrm{t\bar{t}}$+tW sample with diagram removal (DR) in red, the $\mathrm{t\bar{t}}$+tW sample with diagram subtraction (DS) in green, and the bb4l sample in blue, generated with Powheg+Pythia8. [Right] Comparison of the standard $\mathrm{t\bar{t}}$+tW sample with DR in red, the bb4l sample in blue, and the $\mathrm{t\bar{t}}$+tW sample with DR and a higher value of $h_{damp}$ in violet, generated with Powheg+Pythia8. [Updated results have been published in~\cite{ATL-PHYS-PUB-2021-042}.] }
\label{fig:bb4l}
\end{figure}

The so-called \emph{bb4l} sample is a $\mathrm{t\bar{t}}$ sample including all the off-shell effects: double, single and non-resonant contributions. 
This sample improves the description of the off-shell phase space (currently modelled by $\mathrm{t\bar{t}}$+tW) 
for searches and provides a theoretically more solid definition of the top quark mass. 
This sample is also one of the best possible MC setups for $\mathrm{t\bar{t}}$, but it is currently implemented only for different flavour leptons processes, 
making it difficult to use directly in comparisons with data. \\ 
New results from ATLAS are shown in Figure~\ref{fig:bb4l}. 

\section{Conclusions}
ATLAS and CMS have different modelling uncertainties prescriptions, thus a common MC sample would be useful to reduce modelling uncertainties and facilitate ATLAS-CMS combinations. 
A first MC sample (\emph{v0.1}) with Common Settings not yet optimised for agreement with data was successfully produced in both experiments~\cite{ATLAS:2021jgj}. 
The production of the first “physical” common sample (\emph{v0.2}) with settings more tuned to data is ongoing, using common LHE files showered separately in both experiments. \\
The ultimate goal is to produce a real common sample using identical events, with a common Pythia8 tuning using ATLAS and CMS data, and to share resources and prescriptions for nominal and systematic uncertainties.

\bibliography{TOP2021proceedings_Negro}{}
\bibliographystyle{unsrt}
 
\end{document}

%% file: econfmacros.tex
\def\beq{\begin{equation}}
\def\eeq#1{\label{#1}\end{equation}}
\def\eeqn{\end{equation}}

\def\beqa{\begin{eqnarray}}
\def\eeqa#1{\label{#1}\end{eqnarray}}
\def\eeqan{\end{eqnarray}}

\let\bar=\overbar

\def\Dslash{\not{\hbox{\kern-4pt $D$}}}
\def\dslash{\not{\hbox{\kern-2pt $\del$}}}

\def\msb{{\bar{\ssstyle M \kern -1pt S}}}